# Design and development of a near-IR integral field spectrograph for the HWO Coronagraph Instrument

Stephen P. Todd[a *], Dan Dicken[a], Raziye Artan[a], Beth A. Biller[b], Cassandra Mercury[a], Katherine Morris[a], Vinooja Thurairethinam[a], Feng Zhao[c]

[a]UKATC, Royal Observatory Edinburgh, Blackford Hill, Edinburgh, EH9 3HJ, UK; [b]Institute for Astronomy, University of Edinburgh, Royal Observatory Edinburgh, Blackford Hill, Edinburgh, EH9 3HJ, UK; [c]Jet Propulsion Laboratory, California Institute of Technology, USA

## ABSTRACT

The primary mission of the Habitable Worlds Observatory (HWO) is to identify and characterise potentially habitable worlds. Spectra across a wide wavelength range are needed to cover multiple spectral features per molecule of interest. An integral field spectrometer (IFS), fed by a coronograph system, can be used to measure spectra from any planets within the nulled field of the coronograph, while also characterizing the residual speckles as a function of wavelength, enabling the contrast ratio to be further enhanced. We present design trades for an infrared IFS (0.8 to 1.7 μm) for the HWO Coronagraph Instrument, including assessment of the relative merits of lenslet and image slicer based architectures. Key requirements include full sampling of the speckle field at all wavelengths, maximized optical throughput, and control of spectral cross talk and stray light. We identify technology developments needed to advance the instrument design to the required technology readiness level.

## 1. INTRODUCTION

Determining whether rocky exoplanets in nearby habitable zones host life is a central scientific objective highlighted by both ESA Voyage 2050 and NASA's 2020 Decadal Review. While current observatories are largely limited to studying rocky planets orbiting M dwarf stars, where high energy stellar radiation may challenge long term habitability [1]. the most promising targets for discovering life as we understand it are Earth like planets orbiting Sun-like FGK stars. HWO is explicitly designed to observe this population. Operating from a stable L2 orbit with an ultraviolet optical and near infrared platform and a coronagraph capable of achieving star planet contrast levels of order of 10 billion to 1 [2], HWO will be able to detect and spectroscopically characterise the faint reflected light from terrestrial planets around FGK stars. This capability will enable the first robust atmospheric studies of truly Earth analogue worlds.

A key scientific driver for the spectrograph concept is the requirement for broad, simultaneous spectral coverage. There is no single atmospheric feature that provides an unambiguous biosignature. Instead, the current biosignature framework relies on detecting combinations of molecular species that should not coexist in thermochemical equilibrium, such as oxygen, ozone, water vapour, methane, and carbon dioxide [3]. Accessing these diagnostics requires spectroscopy spanning approximately 0.3 to 1.7 microns, ensuring that multiple spectral bands per molecule are observed and that the atmospheric context needed to exclude false positives can be established [4][5], The modern Earth spectrum illustrates this clearly, with oxygen and ozone signatures appearing primarily at ultraviolet and optical wavelengths, while water vapour, carbon dioxide, and methane dominate in the near infrared. Furthermore, Earth's atmospheric composition and associated biosignature gases have evolved markedly over the past four billion years, with the strongest biosignature signals located beyond 0.8 microns for much of Earth's inhabited history. This underscores the scientific necessity of extending HWO's coronagraphic spectroscopy well into the near infrared if the full diversity of potentially habitable planet atmospheres is to be captured.

To meet these science goals, the HWO Coronagraph Instrument will require integral field spectrographs (IFS) in both its UV–optical and near-IR arms. These IFS units simultaneously deliver spatially resolved spectra across the field, enabling:

* Correspondence to SPT – stephen.todd@stfc.ac.uk

- effective suppression and characterisation of residual speckle noise from starlight,
- simultaneous observation of multiple planets in the same system, and
- coronagraphic spectroscopy of extended structures, such as circumstellar debris disks.

For the near-IR arm in particular, the spectrograph is indispensable for detecting key biosignature gases, interpreting atmospheric chemistry, and assessing planetary habitability. This channel also provides rich science beyond terrestrial worlds, enabling deep atmospheric characterisation of gas and ice giants, super-Earths, and mini-Neptunes, as well as Solar System small bodies whose environments can be probed at high contrast. The wavelength coverage and spatial -spectral capability build directly on the development of state of the art instruments such as ELT/METIS, whose combination of infrared integral field spectrograph and coronagraph optics demonstrate the maturity and promise of such an approach.

In this paper we outline the key design drivers for the spectrograph, deriving initial estimates for optical requirements. We identify a number of potential approaches to maximise the efficiency of the instrument for the detection and characterization of planets and the trade studies that will be required to make key design decisions. Details of the coronograph system that sits before the IFS are not within the scope of this study.

## 2. BASELINE DESIGN ASSUMPTIONS

Formal requirements for HWO instruments are still to be defined, but a baseline set of assumptions can be taken from the EAC4 (Exploratory Analytic Case 4) study released by NASA. This assumes an unobscured telescope with 6 m diameter primary mirror. The near-IR IFS sits within the coronograph instrument. The input interface to the IFS is a pupil image in a collimated beam after all the coronagraph optics.

### Wavelength range and spectral resolution

The visible and near-IR channels are each divided into a series of sub-bands with an instantaneous bandpass of ~20%. This is defined by the effective wavelength range of each configuration of the coronagraph optics. A baseline definition of these bands is shown in Figure 1. The near-IR channel would cover 4 bands (F5-F8), corresponding to an overall wavelength range of 952 nm to 1700 nm. We extend our defined wavelength range down to 800 nm to allow some flexibility in defining the crossover between the two channels.

The optimum spectral resolution for these measurements is still an open question. For the purposes of this study we assume a spectral resolution of $R = 100$. Some studies suggest that a high spectral resolution is required for unambiguous detection of biosignatures, however this must be balanced against the signal to noise ratio that can be achieved in each spectral resolution element if the light is dispersed over more pixels on the detector.

Spectral resolution is defined as $R = \lambda/d\lambda$, where $d\lambda$ is the spectral width of one spectral resolution element. To estimate the length of each spectrum on the detector we assume R = 100 and 20% bandpass. The number of spectral elements required to cover a bandpass of $\Delta\lambda$ and centred on $\lambda_0$ is given by

$$L_{spec} = \frac{\Delta\lambda}{d\lambda} = \frac{\Delta\lambda}{\lambda_0}\frac{\lambda_0}{d\lambda} = 0.2R.$$

The length of each spectrum $L_{spec}$ is ~20 spectral resolution elements, equivalent to ~40 detector pixels when assuming 2-pixel sampling per resolution element.

### Field of view and spatial resolution

The critical requirement is to fully sample the PSF at all wavelengths, so the sampling is driven by the shortest wavelength – i.e. 2 spaxels (spatial pixels) per FWHM $\approx \lambda/D$. At 900 nm on a 6-m-diameter telescope $\lambda/_D = 0.15$ µrad $=$ 31 mas. The maximum spaxel size is therefore ~15 mas on-sky. Assuming constant sampling, the image at the longest wavelengths will be oversampled by a factor of 2 in each direction, spreading the flux over 4 times as many spaxels and hence 4 times as many detector pixels. This will decrease the sensitivity at the longest wavelengths to some degree, depending on the noise characteristics of the detector. As such, it may be desirable to have different sampling for different bands within the overall wavelength range of the system.

The required field of view is driven by the expected size of the dark hole that can be produced by the coronograph system. A field of view of 40 λ/D has been given as representative.

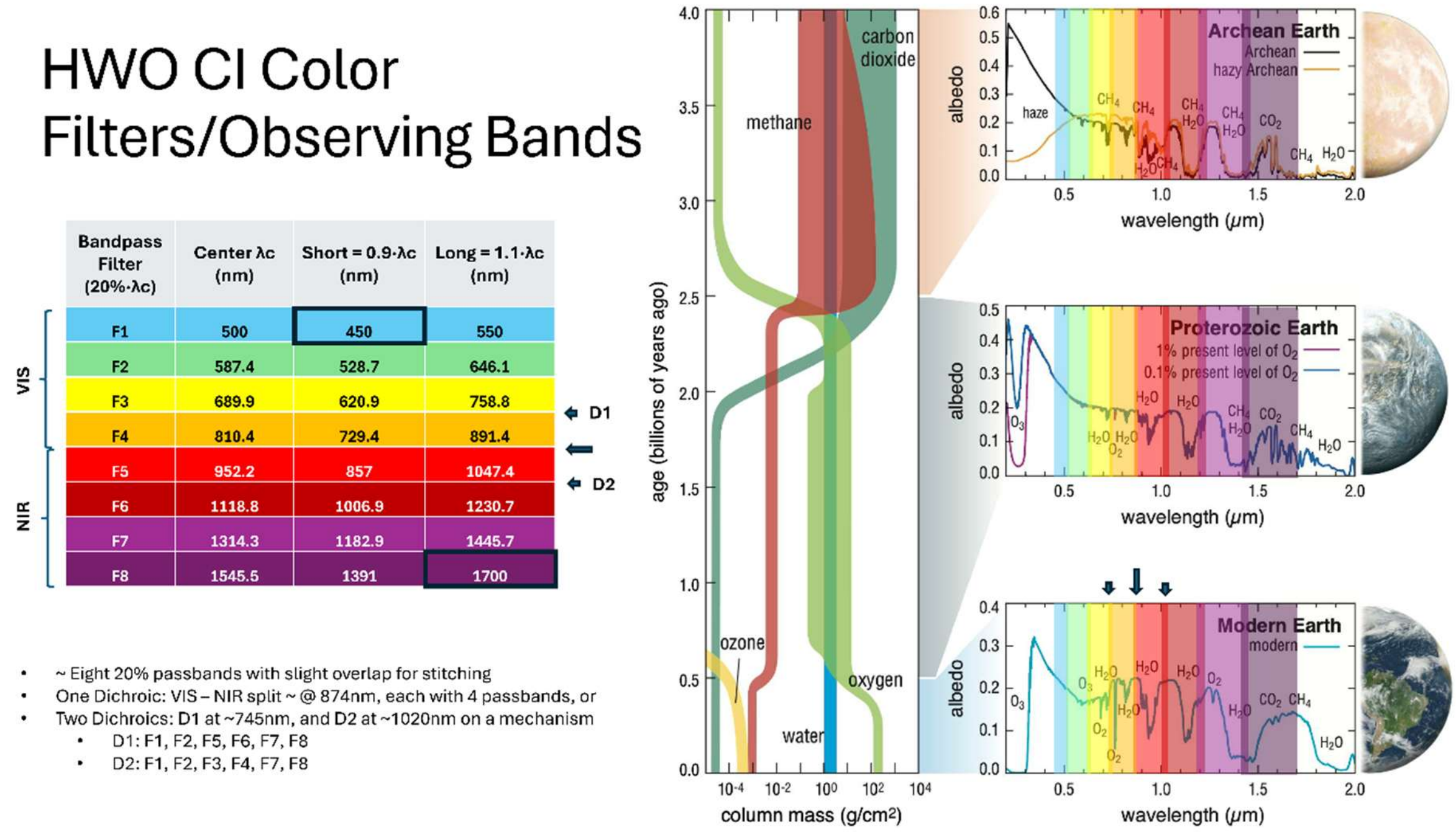

| Bandpass Filter (20%·λc) | Center λc (nm) | Short = 0.9·λc (nm) | Long = 1.1·λc (nm) |
|---|---|---|---|
| F1 | 500 | 450 | 550 |
| F2 | 587.4 | 528.7 | 646.1 |
| F3 | 689.9 | 620.9 | 758.8 |
| F4 | 810.4 | 729.4 | 891.4 |
| F5 | 952.2 | 857 | 1047.4 |
| F6 | 1118.8 | 1006.9 | 1230.7 |
| F7 | 1314.3 | 1182.9 | 1445.7 |
| F8 | 1545.5 | 1391 | 1700 |

Figure 1: The nominal eight coronagraph filter pass bands, comprising four visible and four near infrared bands, overlaid on model spectra of an Earth like planet representing different eons of Earth's inhabited history. The figure illustrates the required wavelength coverage from 0.5 to 1.7 microns to capture the full range of diagnostically important spectral features present across these evolutionary stages. Multiple pass bands are required in each channel because the coronagraph cannot operate efficiently over a single broad wavelength range. *Figure credit: NASA JPL, adapted from Krissansen-Totton et al. [5]*

## 3. IFS ARCHITECTURE

We consider two main architectures for integral field spectroscopy which have both been widely used in astronomy: lenslets and image slicers (Figure 2). Instruments designed specifically for high contrast imaging have generally used lenslets (e.g. GPI, Sphere, Subaru SEEDS IFS), while image slicers have been the technology of choice for more general-purpose instruments, particularly at infrared wavelength (e.g. KMOS, METIS, HARMONI), including proven space heritage on JWST (MIRI and NIRSpec).

The use of fibre bundle based IFUs in combination with lenslet arrays has not been considered in detail here. For this application, there is little advantage from the additional design flexibility over the lenslet approach, while optical throughput will be significantly lower.

**Lenslet based IFU**

The image is focused onto a lenslet array, with each lenslet sampling a single spatial resolution element (spaxel). Each lenslet forms a demagnified pupil image. The resulting sparse grid of micro-pupil images is then used as the entrance focal plane to the spectrograph optics. The dispersion angle and spectral range is set so that the spectra do not overlap and can be separated in data processing. A key benefit of this approach is that the image is sampled at the lenslet focal plane and is therefore insensitive to small optical aberrations in the spectrograph.

The total number of spaxels is constrained as follows

$$N_{pix} \geq \frac{1}{f} N_{spaxels} L_{spec} n_{along} n_{across}$$

Where $f$ is the fill factor on the detector. We assume f = 30% as a starting point. A fully sampled field of view of 40 λ/D requires at least 80 x 80 lenslets ($N_{spaxels} = 6400$). The length of each spectrum in number of spectral resolution elements is assumed to be $L_{spec} = 20$, as described above. $n_{along}, n_{across}$ describe the size of a single spectral resolution element on the detector in pixels (the monochromatic image of a single spaxel), in the dispersion and non-dispersion directions, respectively. As a starting point we assume both are 2 pixels – i.e. a single lenslet forms a 2-pixel diameter spot on the detector. While this is required in the spectral direction, in order to give full sampling of the spectrum, this is undesirable in the cross-spectrum direction since the light is spread over more detector pixels than necessary. Potential solutions to reduce $n_{across}$ closer to 1 pixel are discussed below.

Based on these assumptions, this can be accommodated on a detector with $N_{pix} \geq 2.6 \times 10^6$, and would thus fit onto a 2k-by-2k detector, alongside some margin to either increase the separation of spectra on the detector or to slightly increase the number of spaxels.

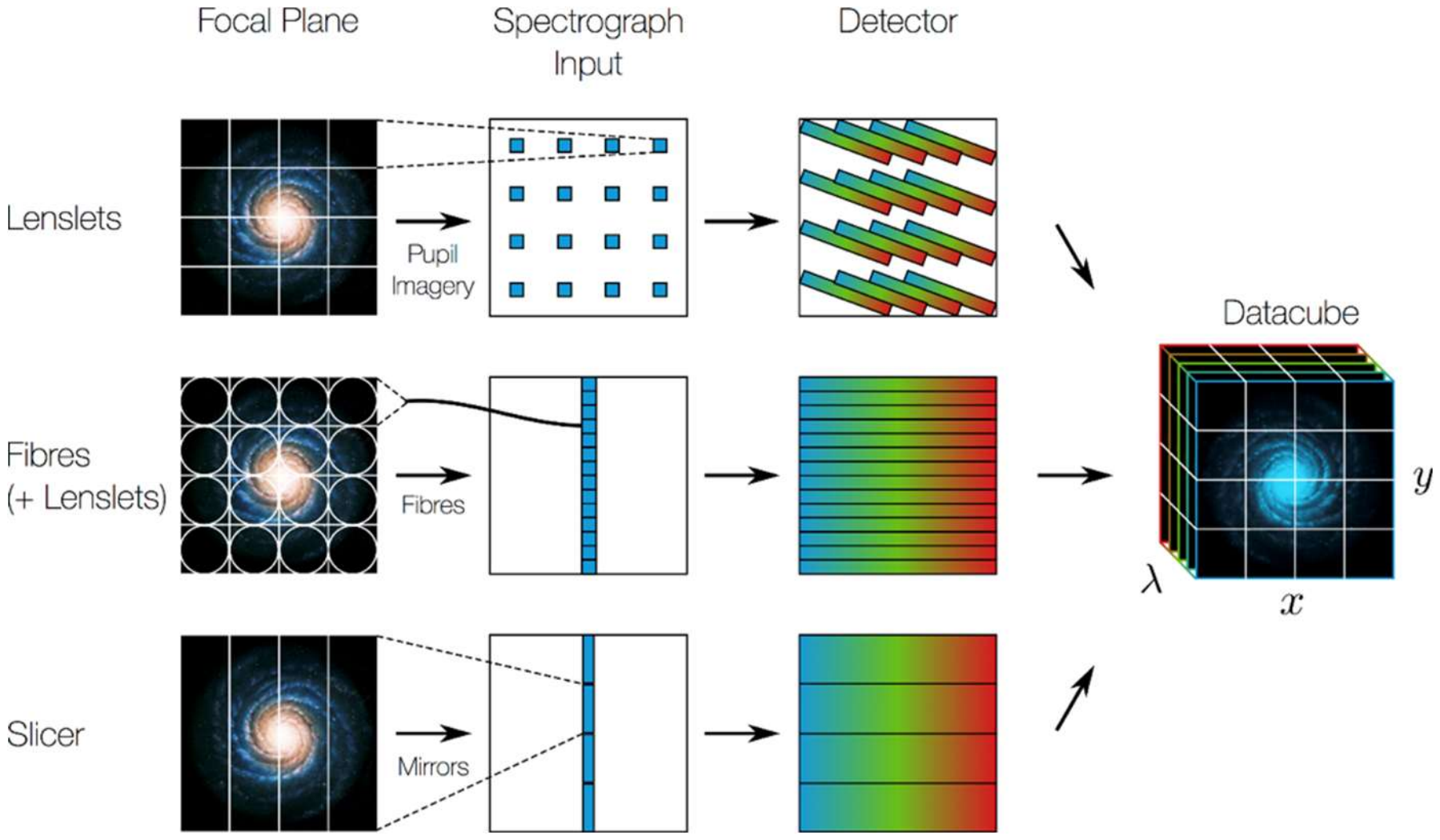

Figure 2: Schematic overview of three common integral field spectrograph architectures used in astronomy. From top to bottom, the figure illustrates lenslet based, fibre coupled, and image slicer approaches. In each case, the focal plane image is spatially sampled and reformatted by the integral field unit before entering the spectrograph, where the light is dispersed and recorded on the detector. The resulting data are reconstructed into a three-dimensional data cube containing two spatial dimensions and one spectral dimension. *Wikimedia Public Domain*

**Image slicing IFU**

The focal plane is imaged onto a slicing mirror, which forms a separate pupil image for each slice of the image. A second multi-facetted mirror array is used to form a line of demagnified slice images which is used as the input slit to the spectrometer.

The image is sampled in one dimension at the slicing mirror. The individual slice images still contain one dimension of spatial information which is sampled by the detector pixels. The total number of spatial resolution elements (spaxels) in the resulting data cube depends on the number of slices ($N_{slice}$) and the length in pixels of each slice image on the detector ($L_{slice}$). If we want to fully sample a square field of view of width 40 $\lambda/D$ this implies that we need $N_{slice} = L_{slice} = 80$. As outlined above, in the spectral direction we need a minimum of 20 spectral resolution elements, with each spectral resolution element sampled by at least 2 detector pixels. We therefore need $2L_{spec} = 40$ pixels in the dispersion direction for each slice.

The width of each slice image on the detector is at least 2 pixels to ensure full sampling of the spectrum. In order to give the same spatial sampling in both dimensions (along and across the slices) the image must be anamorphically magnified before it is projected onto the slicing mirror.The total number of detector pixels required is

$$N_{pix} \geq \frac{1}{f} N_{slice}\, L_{slice} 2L_{spec} = \frac{1}{f}(2.6 \times 10^5)$$

While this seems moderate compared to the capabilities of current IR detectors, there may be a challenge in formatting the spectra in order to use the detector area efficiently. Most image slicing IFS systems generate a single pseudo-slit, containing all of the slice images arranged in one line. On a square detector this fits efficiently if $L_{spec} \approx N_{slice} L_{slice}$ – i.e. when the number of spectral resolution elements is approximately the same as the total number of spaxels. However, this is not the case here. A possible arrangement would be to split the image into 5 columns, with each column containing 16 slice images. This could easily fit onto a 2k-by-2k detector with plenty of margin to avoid cross-talk, as well as potentially allowing for a larger field of view. Moreover, an option for a higher spectral resolution mode could be enabled by adding a focal plane mask somewhere before the image slicer focal plane to illuminate only the central set of 16 slices. This would allow the full width of the detector to be used as the length of the spectrum.

Image slicers require complex optical components. At least two, often three, multi-faceted mirror arrays are required: the slicing mirror itself and mirror arrays to place the output pupil and slice images at the required locations. Monolithic aluminium slicer mirrors and other mirror arrays have been manufactured using single point diamond turning for MIRI, KMOS and other instruments. It would likely be challenging to meet the demanding requirements of HWO through this approach (including surface roughness and form errors). Other IFUs have been assembled from glass slices, polished using conventional techniques to achieve form and roughness tolerances similar to standard glass mirrors. Significant development and testing would be required to space-qualify an assembly of this type. Potential newer technologies that could support the use of glass slices would include the laser inscription approach, giving more freedom in the shape of glass slicing mirrors [6], and the use of hydroxide catalysis bonding to form effectively monolithic components as developed for the LISA project [7].

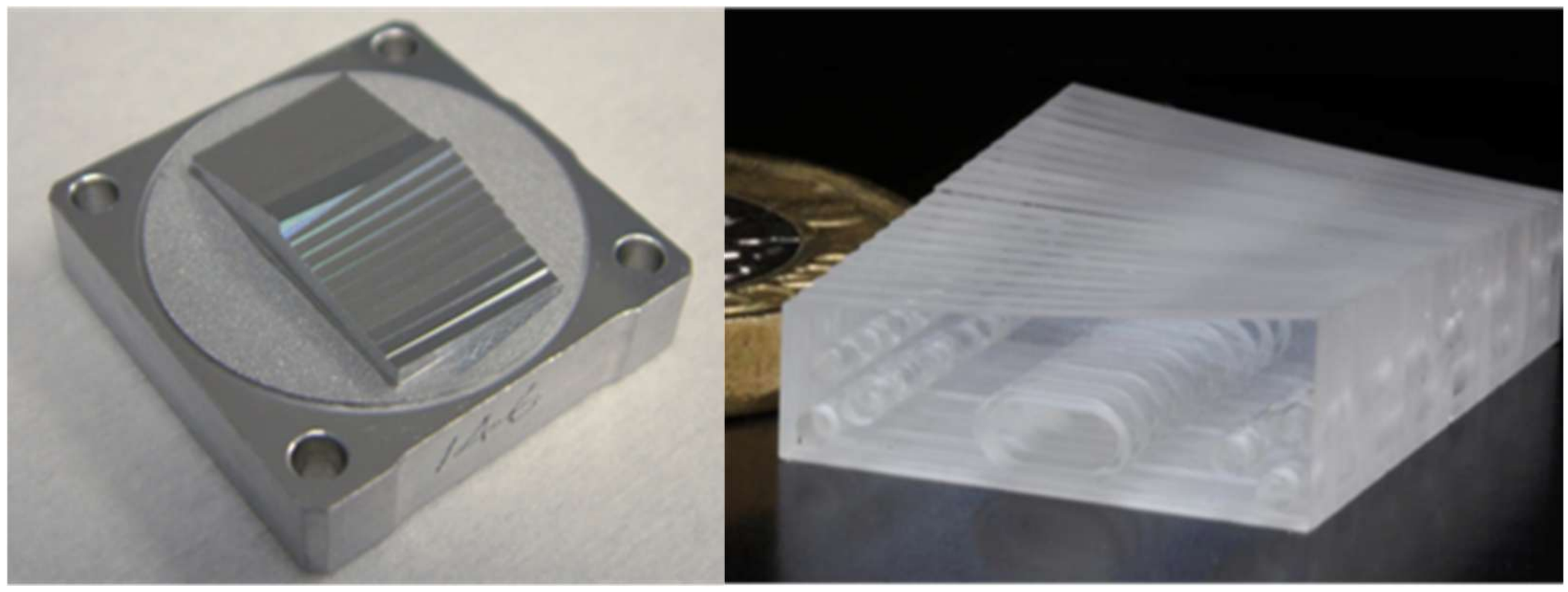

Figure 3: Left - a monolithic aluminium slicing mirror for JWST MIRI. Right - a prototype glass slicer made using ultra-fast laser assisted etching (made at Institute of Photonics and Quantum Sciences, Heriot-Watt University, Edinburgh)

**Mirrorlets**

A hybrid approach of using 'mirrorlets' has been proposed to combine advantages of the lenslet and image slicing approach [8]. The image is sampled by lenslets or mirrorlets at the focal plane, but these are grouped to produce slitlet images similar to the output of a slicing system. The primary benefit of this approach is to achieve a high packing efficiency of spectra on the detector while maintaining many of the advantages of the lenslet approach. This is not considered in further detail within this study but should be assessed in future design trades.

**Baseline IFS architecture selection**

The lenslet approach samples the image in both dimensions at the lenslet array. The spatial image quality of the data is insensitive to small optical aberrations in the spectrograph optics after the lenslet array. The number of optical surfaces will almost certainly be lower for the lenslet based approach, resulting in higher optical throughput.

In an image slicing system, on the other hand, the sampling in one dimension occurs at the detector and as such, the image quality budget must be allocated across the full system to the detector focal plane. Additionally, anamorphic relay optics are required to give the required spatial and spectral sampling in an image slicing system. The main advantage of image slicers is that they allow more efficient packing of the spectra onto the detector array, so a larger field of view and/or spectral range can be accommodated on a given size of detector. However, this does not appear to be a driving requirement in this application.

At this stage we do not rule out the image slicing approach, but within the scope of this study we focus on understanding the performance and key design trades of the lenslet approach.

## 4. BASELINE OPTICAL DESIGN CONCEPT

An initial optical design concept was created to give a baseline. This is not proposed as a design for the real system, but similarly to the NASA Exploratory Analytic Cases is intended to give a starting point for analysis in order to identify potential areas to improve performance.

The initial concept, shown in Figure 4, has been developed based on the design of the Prototype Imaging Spectrograph for Coronagraphic Exoplanet Studies (PISCES) [9], which was designed to operate in the visible range. Light enters the spectrograph through the array of lenslets. The design includes a refractive collimator and camera. Dispersion is provided by a compound prism, composed of a fused silica prism and a ZnS compensator plate – a shallow prism wedged in the opposite direction. This combination of materials compensates for variation in dispersion with wavelength, reducing the variation in spectral resolution across the wide wavelength range of the NIR arm.

Both the collimator and camera doublets use $CaF_2$, $BaF_2$ and fused silica lenses, which have been shown to be relatively radiation-resistant, provided that the materials are of high purity (see e.g. [10][11]). This conceptual design demonstrates that it is feasible to employ relatively simple spectrometer design to achieve diffraction limited image quality. The use of reflective optics for collimator or camera has not been ruled out and will be considered further in a future phase of the study.

Analysis of the optical design concept identified a number of potential approaches to improve the performance. These will form the basis of optical trade studies to be completed as part of the ongoing phase of the design process as discussed below.

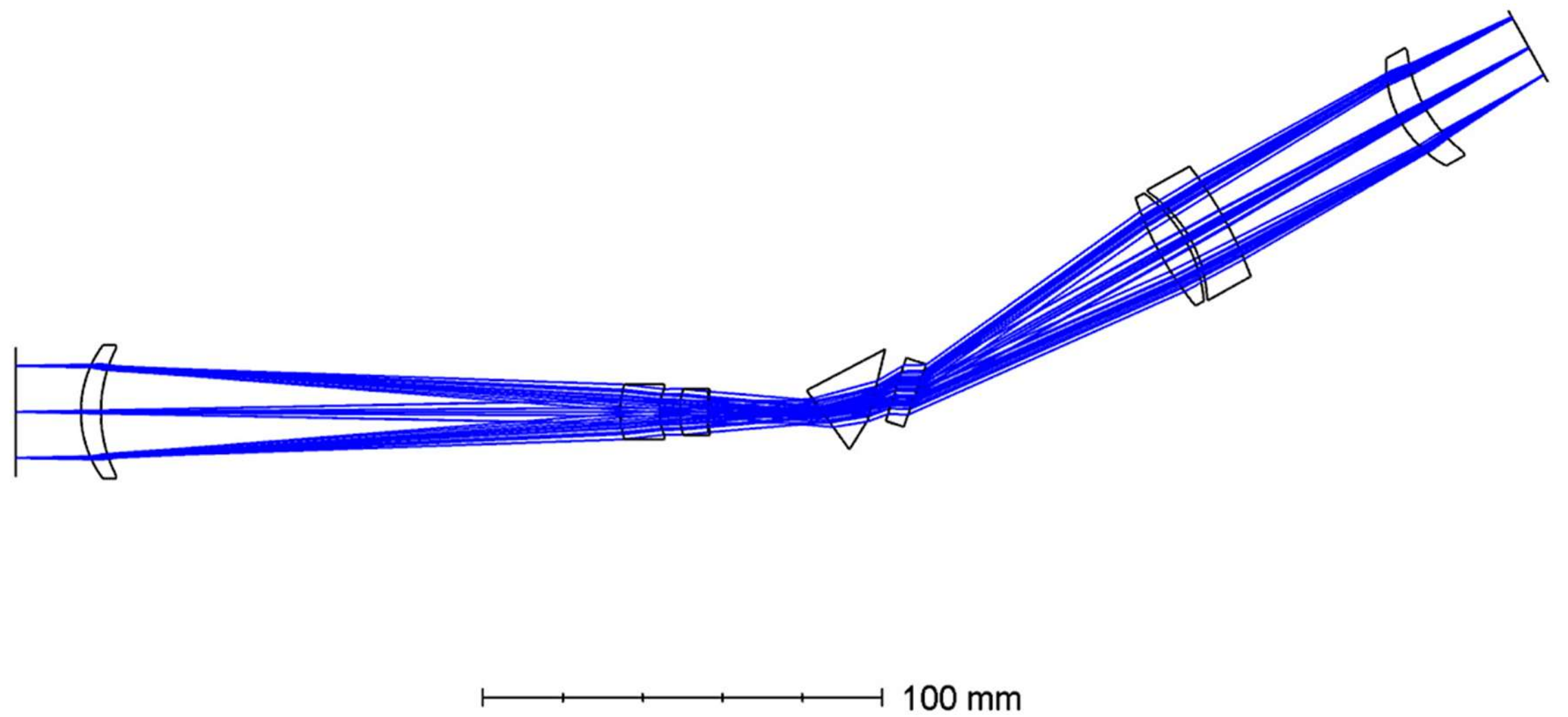

Figure 4: Conceptual design of a HWO NIR integral field spectrometer, excluding the lenslets / slicer and relay optics.

## 5. MINIMISING CROSS-TALK

All lenslet based integral field spectrographs will see some degree of cross-talk between adjacent spectra due to the wings of the re-imaged micropupils as a result of diffraction. While this can be compensated to some degree during data processing, it is desirable to minimise this effect in the design itself. A mask, consisting of an array of pinholes, is generally placed at the micro-pupil plane. This may be applied directly to the reverse surface of the lenslet array.

An alternative dual lenslet approach, known as BIGRE is used in the SPHERE instrument on ESO VLT. A second lenslet array is used, potentially on the reverse of the same component, forming a relatively thick double-sided lenslet array, as illustrated in Figure 5. This second array then forms demagnified images of each input lenslet. In this case, the input to the spectrometer will be a grid of demagnified images rather than pupil images. A pupil mask within the spectrometer optics forms an integral part of the system, acting as a spatial filter. The size of this pupil mask can be optimised to give a spot profile with wings that decay more rapidly than the airy pattern that would be seen from the re-imaged micropupils.

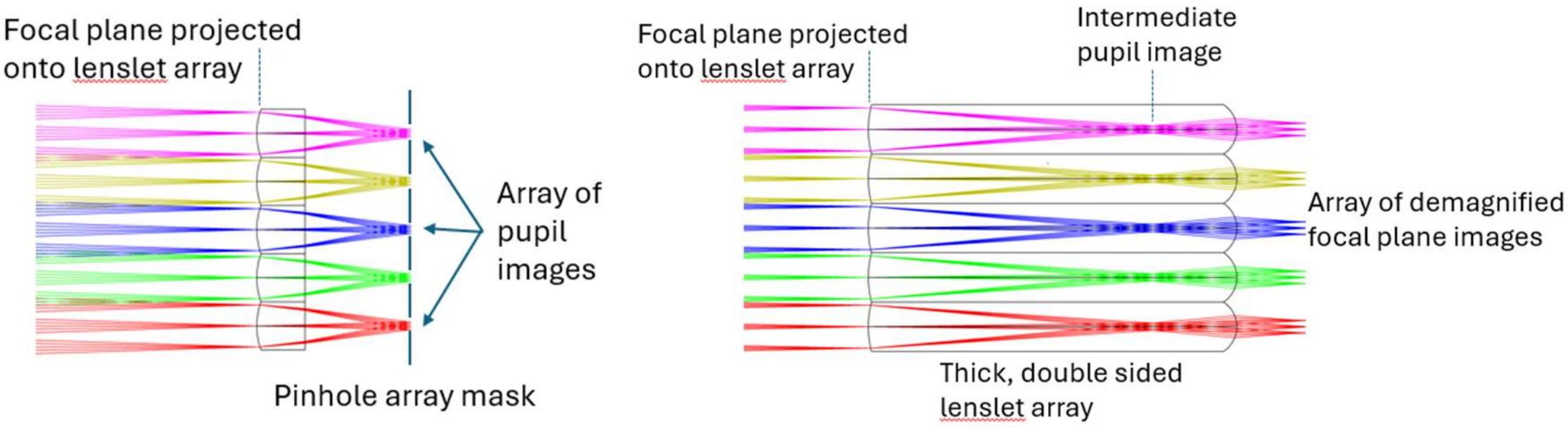

Figure 5: Left – the approach used in many IFS instruments with a pinhole mask at the micropupil array. Right – the BIGRE approach with a second lenslet array used to form demagnified images of the input lenslets.

## 6. ACHIEVING GOOD SAMPLING AT ALL WAVELENGTHS

For high contrast imaging applications, it is essential to fully sample the image at all wavelengths, generally taken as 2 spaxels across a diffraction limited spot size of λ/D. The sampling is therefore set by the shortest wavelength of the instrument. At the longest wavelengths this would result in up to 4 spaxels across a diffraction limited spot, oversampling the image by a factor of 2 in each direction. This may result in a negative impact on the signal to noise ratio, depending on the noise properties of the detectors. The yield of planets detected at 1.6 µm is expected to be 1/4 to 1/3 of the yield at 1.0 µm [13]. Maintaining good sampling at the long wavelengths will therefore aid in maximising the yield across all wavelengths.

There is a trade study to be done, balancing the extra complexity of switching sampling scales for different wavelength ranges against the potential performance gain. In the discussion below we consider options for switching between two sampling scales, one for the two shorter wavelength bands (0.9 – 1.30 µm) and another for the longer wavelength bands (1.25 – 1.70 µm). This reduces the maximum oversampling from a factor of 1.9 down to 1.4. The same principle could be applied to implement 3 or 4 different sampling scales if required.

One potential approach to achieve sufficient sampling across the NIR waveband is the employment of interchangeable pre-optics to provide different image scales for the longer and shorter wavebands (Figure 6). A mechanism would be used to move different lens barrels or sets of mirrors into the beam between the entrance pupil and the lenslet array. Each set of relay optics would need to produce a focal plane at the same location with pupil image at infinity (telecentric). The lenslets remain fixed with respect to the spectrograph optics and detector, so positioning errors in the mechanism do not shift the spectra on the detector, but appear as a pointing error, shifting the whole image relative to the lenslets. The image is highly magnified at the lenslets, so a shift of 15 µm corresponds to 0.1 spaxels, or 1.5 mas on sky.

An alternative solution would be an interchangeable microlens array mechanism (Figure 7). The lenslet sizes should be matched to λ/2D at the shortest wavelength within each sub-band. The focal length of each individual lenslet would remain the same, so that a larger numerical aperture must be accepted by the spectrograph optics. The magnification from the lenslet array to the detector remains unchanged, and so the spacing between spots is increased. The number of spaxels that can be measured on a given detector area is reduced. The field of view measured on-sky remains the same given the coarser sampling.

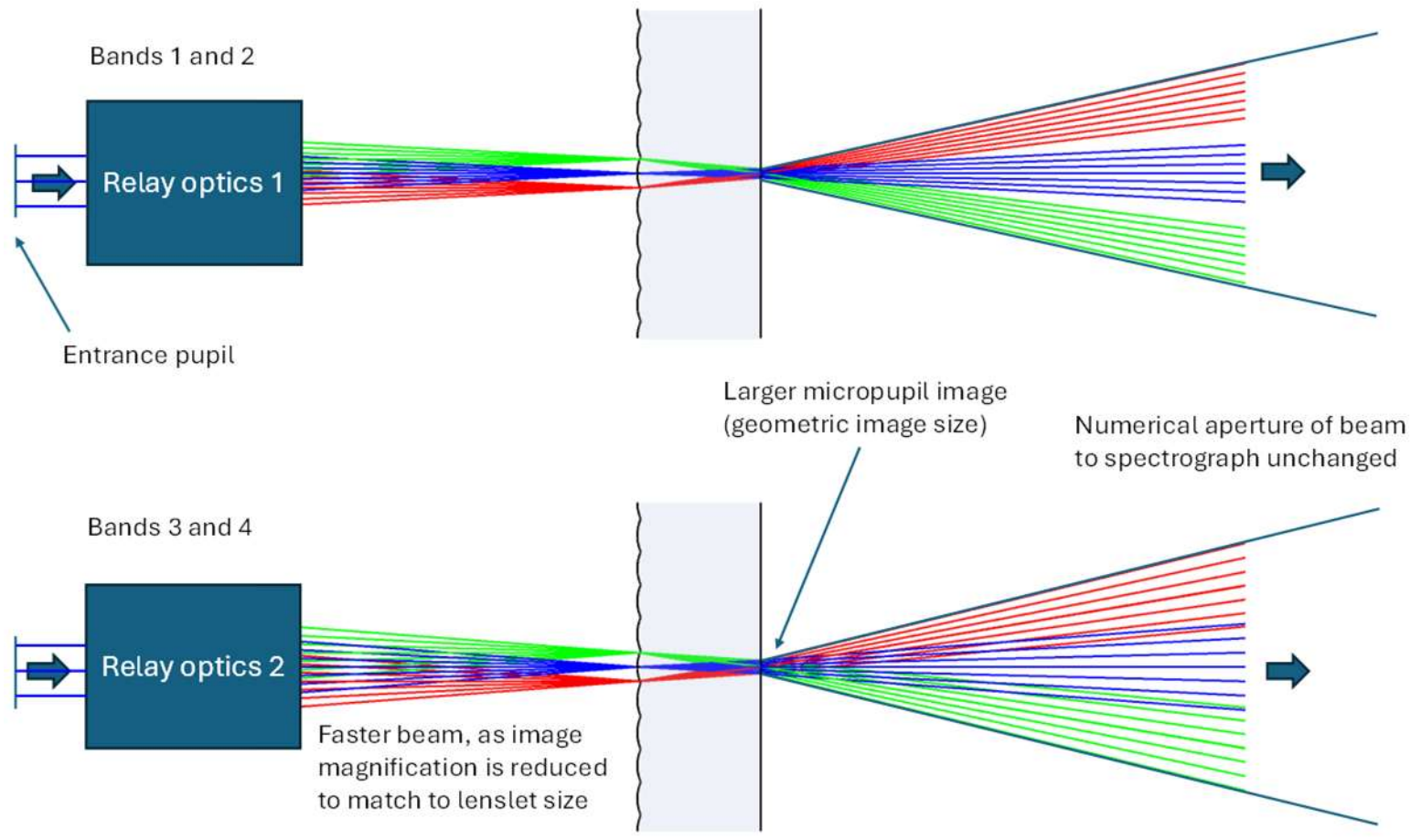

Figure 6: Changing the pre-lenslet relay optics to give different sampling at different wavelengths (not to scale).

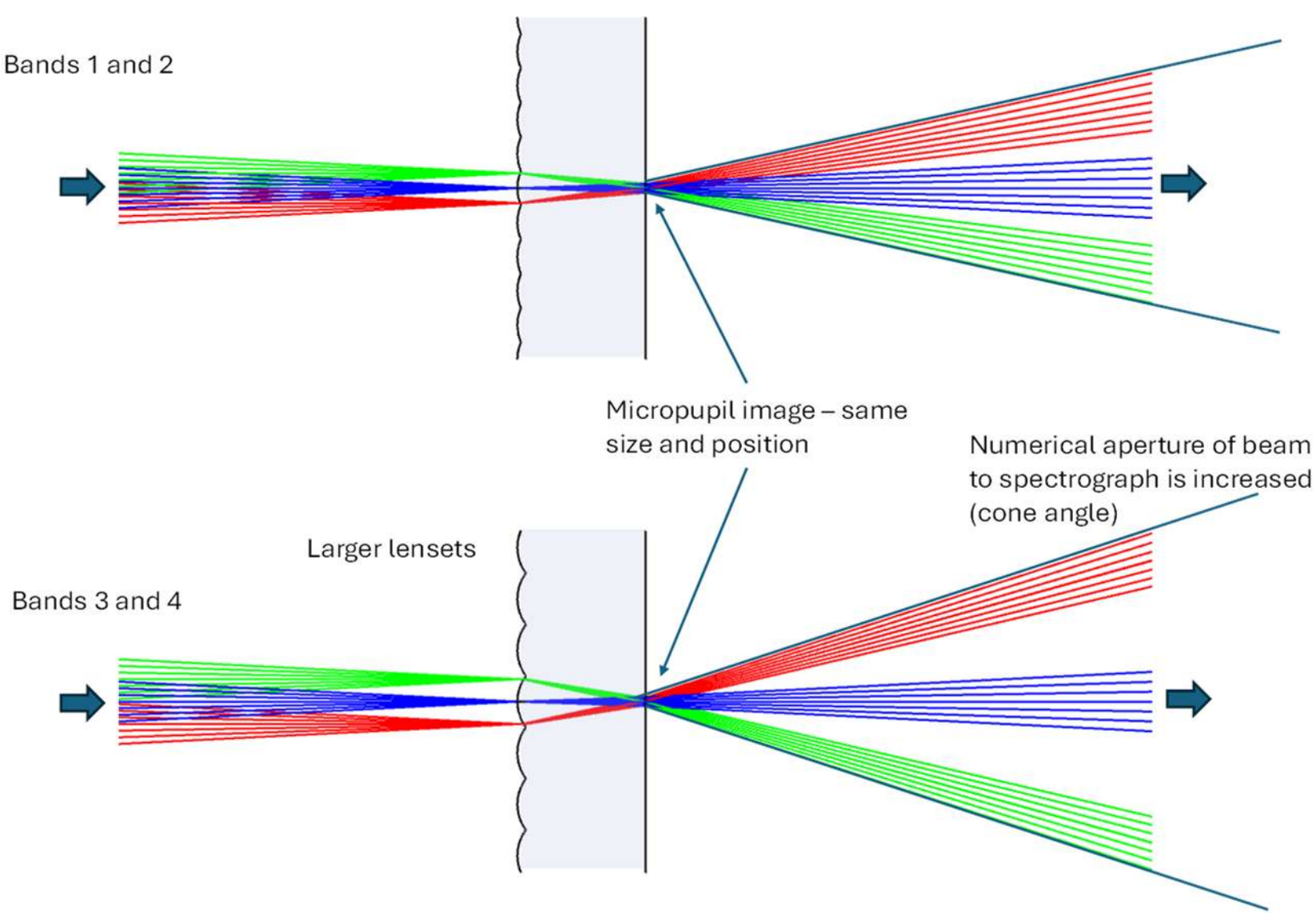

Figure 7: Changing lenslet array to give coarser sampling at long wavelengths. Different coloured rays indicate field points at the centre and edges of a single lenslet. (Not to scale)

A trade study is required to assess three options:

1. Single spaxel scale across the entire NIR IFS wavelength range
2. Interchangeable lenslet arrays
3. Interchangeable relay optics to change the scale at the lenslet array

Further modelling is needed to fully understand the implications of approaches (2) and (3) with regard to switching scales. In particular, the effects of diffraction and spatial filtering must be modelled, including the effects of spectrograph optical apertures. Experience on HARMONI [14] and multiple papers (including [12]) have shown that these effects can be counter-intuitive in IFS instruments operating at the diffraction limit and have significant impact on performance.

Mechanically, exchanging lenslet arrays only requires a single small component to be moved, but to a high level of precision. Deploying different sets of relay optics would need a larger mechanism, but the positioning accuracy requirements are likely at least an order of magnitude less stringent.

The expected performance gain will depend on the noise properties of the detectors. If the read noise of the detectors is sufficiently low, then the oversampled signal can be re-binned with little additional noise.

## 7. INTRODUCING ANAMORPHIC MAGNIFICATION

In order to achieve full sampling of the spectrum, the spot produced by each lenslet must be imaged to have a FWHM of at least 2 pixels on the detector. This also means that each spectrum is spread over multiple pixels in the across-spectrum direction. Extracting the spectrum will include read noise from all of these pixels. Consequently, to maximise the signal to noise, we aim to concentrate the spectrum onto as few pixels as possible in the non-dispersion direction. To do this whilst maintaining equal spatial sampling in both dimensions, requires anamorphic magnification to be introduced. Ideally, we intend for the spot on the detector, produced by a single lenslet illuminated by a monochromatic source, to be 2 pixels wide in the dispersion direction and 1 pixel wide in the perpendicular direction. There are multiple possible approaches which should be considered as part of a more detailed design study.

### Prism geometry

Prisms can introduce anamorphic magnification in the beam as a refractive effect due to a difference in the beam angles at the input and output surfaces of the prism. If this anamorphism is introduced in the dispersion direction, it may be possible to exploit this effect of the prism geometry. The prism input angle and its apex angles can thus be adjusted to provide asymmetrical magnification in the dispersion and non-dispersion directions. This effect is unlikely to provide the factor of 2 that we are aiming for but can be set to make a contribution to the required anamorphism.

### Anamorphic collimator or camera optics

Anamorphic magnification can be introduced in the collimator or camera optics through the use of toroidal surfaces. The axes of the anamorphic image can be set parallel to the dispersion direction as illustrated in Figure 8.

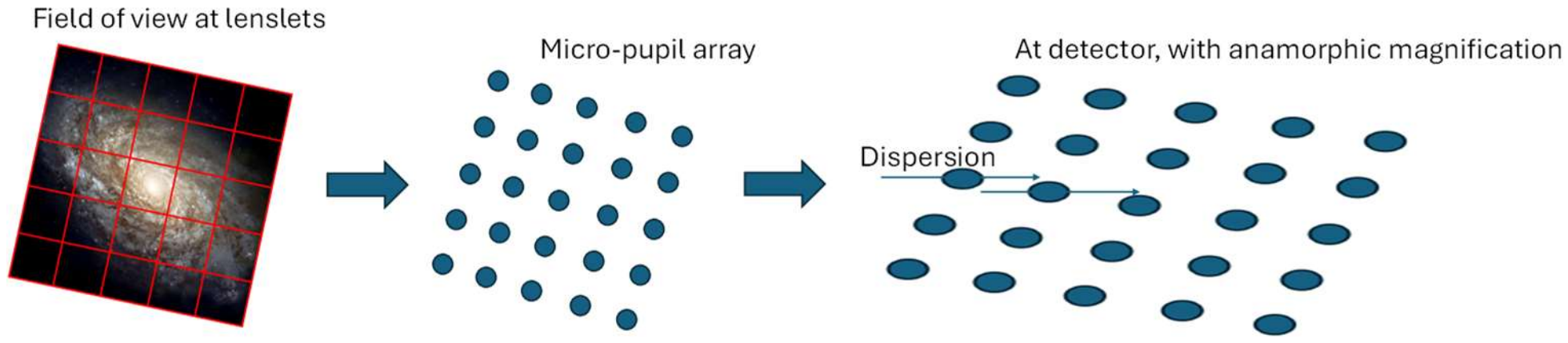

Figure 8: Anamorphic magnification of the full IFS field using the collimator or camera

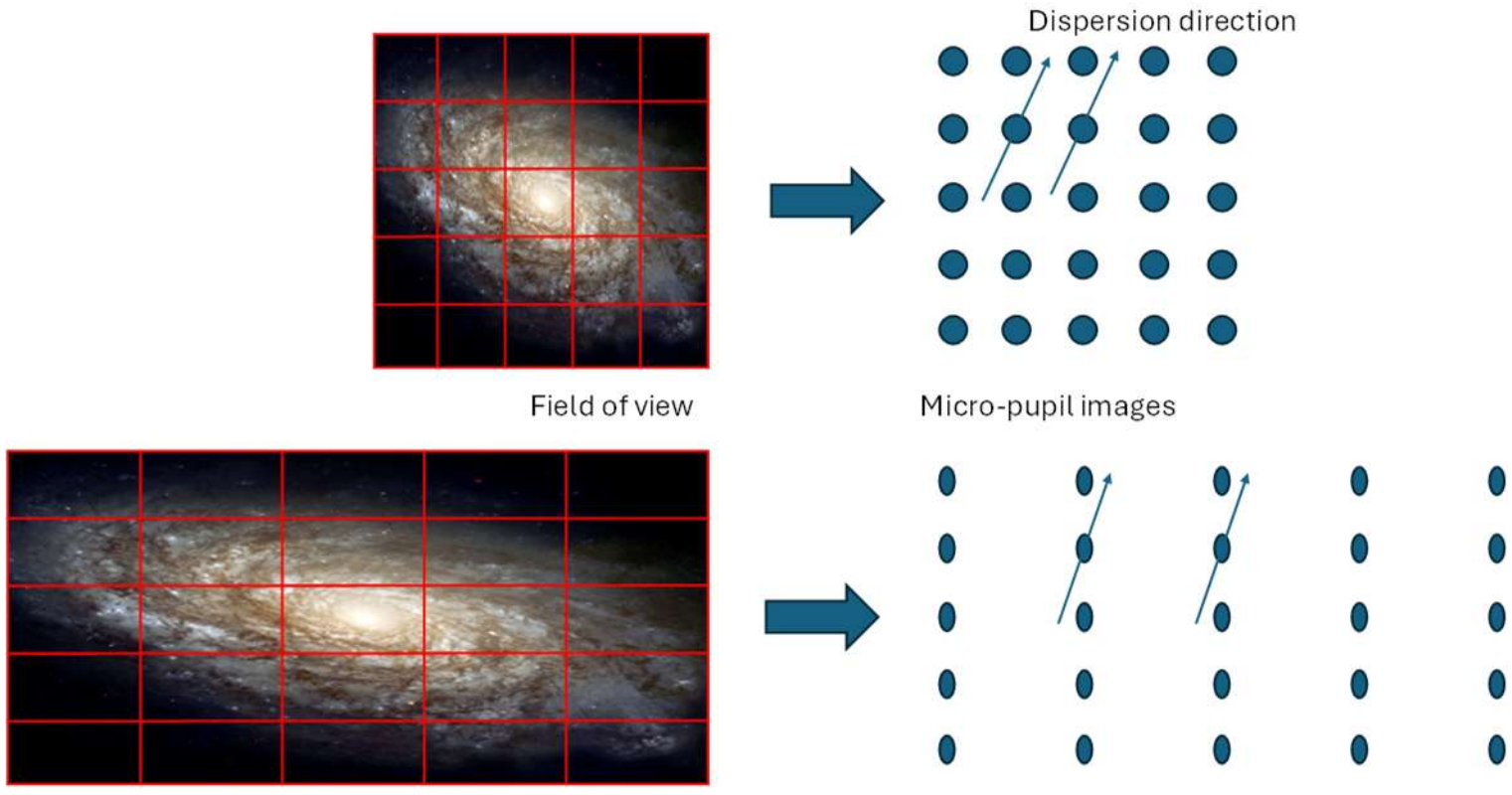

Figure 9: A schematic view of the effect of introducing anamorphic magnification before the lenslet array.

**Anamorphic relay optics**

Anamorphic magnification in the relay optics before the lenslet array can similarly be used to stretch the image in one direction. Equal sampling in both directions can be maintained by using rectangular lenslets, while the surface of the lenslets remains spherical. Magnifying the image in one dimension will give a slower focal ratio and hence a smaller pupil image in this same direction as illustrated in Figure 9. The axes of the anamorphic magnification must be aligned to the axes of the lenslet array, so the dispersion direction cannot be exactly aligned to the long axis of the micropupil images, but a significant improvement could still be achieved.

**Anamorphic lenslet array**

An alternative approach would be to use the lenslets themselves. A lenslet array with toroidal facets on the front surface and cylindrical facets on the rear surface can directly generate elliptical spots with no additional optical surfaces in the path. Since the anamorphic factor is applied to the field of each lenslet individually, the resulting array of micropupil images is still arranged on a square grid. Further work is required to understand the manufacturing implications, including an assessment of the potential for additive manufacture using the 2-photon polymerisation process.

Minimising the number of detector pixels for each spaxel has been identified as one of the priority areas for maximising the yield of detected exoplanets. A trade study will be carried out to select the most promising approach, including:

- Optical modelling – do all of these schemes provide equivalent benefits when diffraction of the micro-pupil images is fully considered?
- Manufacturing and alignment constraints – all options involve more complex, non-circularly symmetric optics.
- Compatibility with interchangeable image sampling implementation – the other area which is desirable to maximise signal to noise, as outlined in Section 6.
- Technology status, particularly with regards to manufacturing capabilities of anamorphic lenslet arrays.

## 8. NEXT STEPS

In this initial phase of the design study we have outlined the parameters driving the design of the near-IR IFS for HWO. In the process we have identified a number of key design decisions to optimize the performance of the instrument, maximising the potential yield of characterised earth-like planets. We have defined a series of focussed optical design studies which will allow us to move towards a well-developed conceptual baseline design. These include:

- Minimising cross-talk between lenslets
- Optimising the spatial sampling at all wavelengths
- Using anamorphic magnification to concentrate the spectrum onto fewer detector pixels.

Alongside this we will continue to evaluate and prototype key technologies, including novel processes for manufacturing lenslet arrays such as additive manufacture.

The HWO coronagraph instrument places exceptionally stringent requirements on its near infrared focal plane arrays. Detectors must be capable of registering individual photons across a large field of view while introducing effectively no additional instrumental noise. At the extremely low photon fluxes relevant to exoplanet coronagraphy, every detected photon carries scientific value, and detector performance directly constrains achievable contrast, sensitivity, and overall observing efficiency. A good understanding of the expected detector performance is required to fully carry out the optical trade studies outlined above. We are working to develop an ultra-low background test facility to characterise devices such as the Leonardo APD (avalanche photodiode) detectors in the photon counting regime.

The design studies so far have assumed a 6 m telescope diameter and detector arrays with 2k x 2k pixels. As we move forwards we will consider the implications of the NASA EAC 4 and EAC 5 concepts for the observatory, including primary mirror diameters of 8 – 10 m. We will also assess the potential performance gains that would be enabled by a 4k x 4k detector array to build a case to support future development of these devices.